\documentclass[10pt,prl,twocolumn,superscriptaddress]{revtex4}

\usepackage{graphicx}
\usepackage{amssymb}
\usepackage{times}
\usepackage{amsmath}
\usepackage{amsfonts}
\usepackage{txfonts}
\usepackage{color}

\begin{document}

\title{Cooling and control of a cavity opto-electromechanical system}

\author{Kwan H. Lee} \affiliation{Department of Physics, University of Queensland, St Lucia,
Queensland 4072, Australia}

\author{Terry G. McRae} \affiliation{Department of Physics, University of Queensland, St Lucia,
Queensland 4072, Australia} \affiliation{MacDiarmid Institute, Physics Department, University of Otago, Dunedin, New Zealand}

\author{Glen I. Harris} \affiliation{Department of Physics, University of Queensland, St Lucia,
Queensland 4072, Australia}

\author{Joachim Knittel} \affiliation{Department of Physics, University of Queensland, St Lucia,
Queensland 4072, Australia}

\author{Warwick P. Bowen} \affiliation{Department of Physics, University of Queensland, St Lucia,
Queensland 4072, Australia}

\begin{abstract}
We implement a cavity opto-electromechanical system integrating electrical actuation capabilities 
of nanoelectromechanical devices with ultrasensitive mechanical transduction achieved via intra-cavity opto-mechanical coupling.  Electrical gradient forces as large as 0.40~$\mu$N are realized, with simultaneous mechanical transduction sensitivity of $1.5 \! \times \! 10^{-18}$~m~Hz$^{-1/2}$ representing a three orders of magnitude improvement over any nanoelectromechanical system to date.
Opto-electromechanical feedback cooling is demonstrated, exhibiting strong squashing of the in-loop transduction signal. Out-of-loop transduction provides accurate temperature calibration even in the critical paradigm where measurement backaction induces opto-mechanical correlations.
\end{abstract}


\date{\today} \maketitle

Mechanical oscillators are predicted to exhibit striking quantum behavior\cite{Schwab05}; enabling experimental tests of long-standing scientific problems such as quantum gravity\cite{Bouwmeester03} and quantum nonlinear dynamics\cite{Rugar2,Milburn,Katz}, as well as far-reaching applications in metrology\cite{Rugar} and quantum information systems\cite{Mancini}. Rapid progress towards this quantum regime is underway in both cavity optomechanical systems (COMS)\cite{Cleland09} and nanoelectromechanical systems (NEMS)\cite{Lehnert,Schwab}. COMS enable ultrasensitive transduction of the mechanical motion, presenting a solution to the key challenge of resolving the oscillators quantum zero-point fluctuations\cite{Kippenberg09}. To date, however, mechanical  actuation in COMS has been achieved via radiation pressure\cite{Cohadon,Kleckner,Kippenberg09}, which is inherently weak and severely constrained in the quantum regime by heating from intra-cavity optical absorption\cite{Kippenberg09}.  The electrical actuation of NEMS, by comparison, can be orders of magnitude stronger and is far less prone to heating\cite{Schwab,Poggio,Roukes08}; providing access to nonlinear mechanical behavior\cite{Rugar2}, as well as greater scope for quantum control and cooling\cite{Schwab,Poggio,Unterreithmeier09}. 

Recently, a non-dissipative electrical actuation technique using localized gradient forces has been developed for dielectric NEMS\cite{Unterreithmeier09}.
In this Letter we report a cavity opto-electromechanical system (COEMS) which extends this technique to COMS based on silica microtoroids on a silicon chip. The microtoroid structure integrates high quality optical and mechanical resonances; while the dielectric nature of silica is naturally suited to gradient force actuation\cite{Unterreithmeier09}. Electrical gradient forces as large a 0.40~$\mu$N are achieved, enabling strong mechanical actuation without observable heating effects. Simultaneously, ultrasensitive optical transduction is implemented at the level of $1.5 \! \times \! 10^{-18}$~m~Hz$^{-1/2}$  close to the mechanical zero-point motion and surpassing the current state-of-the-art in NEMS by three orders of magnitude\cite{Knobel}.

\begin{figure}[t!]
\begin{center}
\includegraphics[width=8.5cm]{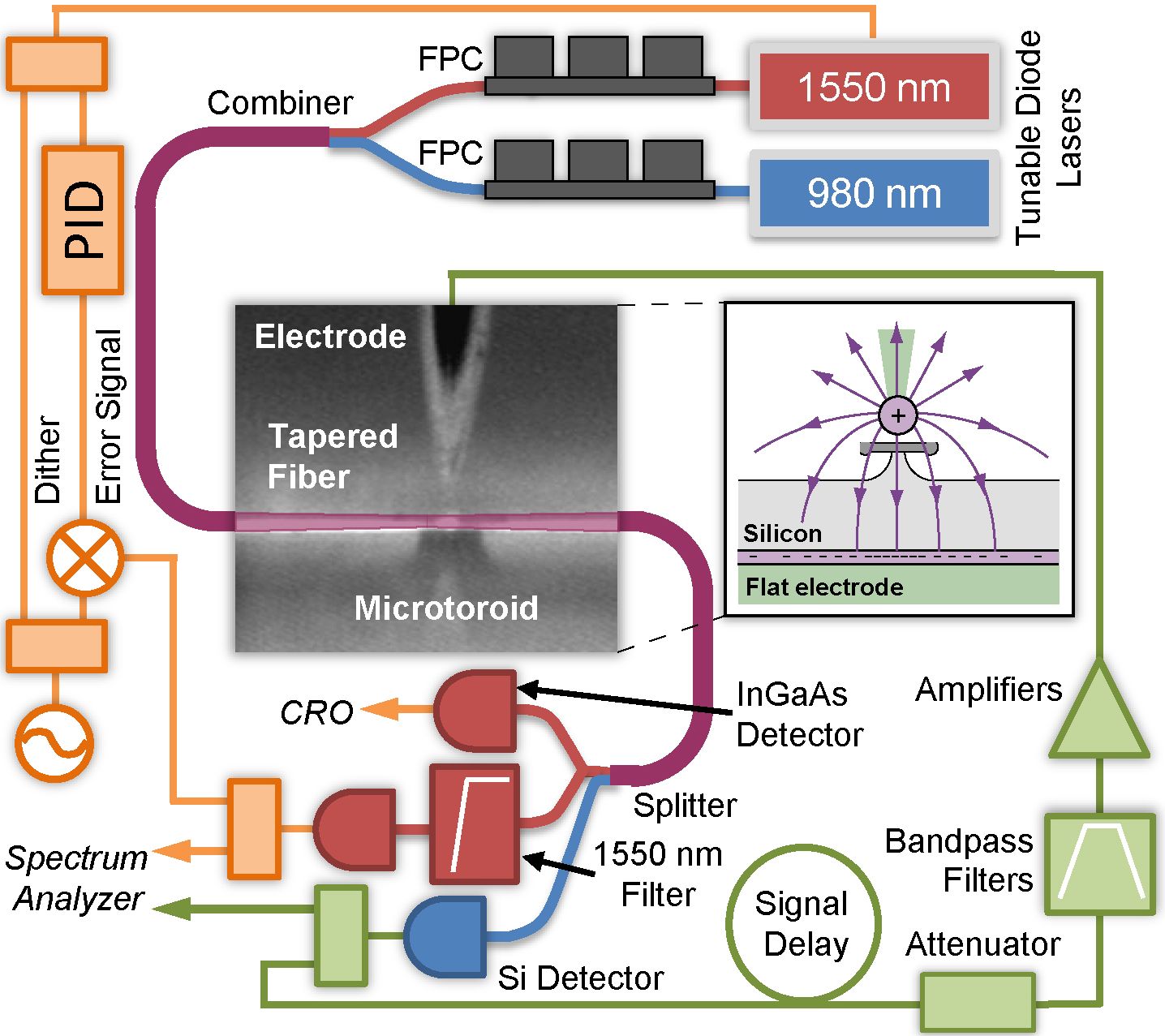}
\caption{(color online) Experimental schematic including electronic locking method for in-loop probe. The out-of-loop probe was thermally locked. FPC: fibre polarization controller. Inset:  electric field distribution between electrodes. All experiments were carried out at room temperature ($T=300$~K) and atmospheric pressure.}
\label{expt}
\end{center}
\end{figure}
Electo-mechanical actuation and opto-mechanical transduction, when combined within a feedback loop, allow immediate control of the state of the mechanical oscillator;  with the capacity to facilitate, for example, feedback cooling or heating\cite{Cohadon}, electro-optic spring effects\cite{Vahala}, and  phonon lasing\cite{Vahala2}. Here, feedback cooling is implemented as a demonstration. All previous COMS and NEMS feedback cooling experiments have used a single in-loop transducer for both cooling and characterization of the mechanical oscillator\cite{Cohadon,Poggio,Schwab}. However, anti-correlations are established between the mechanical motion and the transduction noise, which cause {\it squashing} and an over-estimate of the achieved cooling\cite{Poggio}. Critically, the in-loop transduction signal predicts a homogeneously {\it decreasing} temperature with increasing gain, when in fact transduction noise imprinted on the mechanical motion causes net {\it heating} at high gain.
Here, we implement for the first time a second out-of-loop transducer\cite{Kippenberg09}, providing accurate temperature characterization even in the presence of strong squashing. This allows the first direct observation of the transduction noise limit of feedback cooling. Out-of-loop transduction is important in any circumstance where the mechanical motion becomes correlated to the transduction noise. Particularly critical is the quantum paradigm where such correlations are inherently established
by radiation pressure induced backaction.

The capacity to strongly electrically actuate COMS represents an enabling step towards the experimental realization of mechanical nonlinear dynamics; with nanofabrication techniques providing the means to engineer the nonlinear properties. Such dynamics are traditionally the realm of NEMS, where arrays of coupled mechanical oscillators allow, for example, oscillation synchronization\cite{Cross}, enhanced sensing architechtures\cite{Spletzer}, sub-Heisenberg limit metrology\cite{Milburn}, and mechanical quantum state engineering\cite{Katz}. COMS, however, present the advantage of superior transduction sensitivity, affording the possibility of achieving the new regime of quantum nonlinear dynamics. Furthermore, integration into ultracold superconducting circuits has the potential to unify cavity opto-mechanics and the new and burgeoning field of circuit quantum electrodynamics\cite{Girvin08}.

A schematic of our experimental setup is shown in Fig.~\ref{expt}. To achieve gradient force actuation a radio-frequency (RF) voltage was applied to a sharp
stainless steel electrode with 2~$\mu$m tip diameter positioned 15~$\mu$m vertically above a microtoroid with major
and minor diameters of 60~and~6~$\mu$m, respectively, and a $10~\mu$m undercut. The height was chosen to balance the stronger gradient forces achievable close to the microtoroid, with the potential for chemical contamination and structural damage due to physical contact.
A grounded flat electrode was mounted beneath the silicon chip with the combination of electrodes forming a capacitor. The applied voltage induced a point-like charge build-up on the sharp electrode tip, with a corresponding sheet of opposite charge on the flat electrode, as shown in the inset in Fig.~\ref{expt} and confirmed through finite element modeling. Since the microtoroid was in close vicinity to the sharp electrode, the electric field it experienced was well approximated by that of a point charge. The strong gradient of such a field, combined with surface charge induced static electric fields which polarize the microtoroid, enabled large gradient forces to be exerted.

Widely tunable external cavity diode lasers at 1550~and~980~nm respectively provided in- and out-of-loop optical {\it probe fields}. Both fields were evanescently coupled into a microtoroid using a tapered optical fiber held in close proximity to the microtoroid surface, and were
frequency detuned to the full-width-half-maximum of whispering gallery modes with relatively high coupled optical quality factors of $Q \! \approx \!  10^6$. Mechanical motion altered the cavity's optical path length, and as a result was transferred to the amplitude of the out-coupled in- and out-of-loop optical probes\cite{Carmon05}. The out-coupled probes were split, and detected on Si and InGaAs photodiodes; respectively providing in- and out-of-loop electronic transduction signals.

\begin{figure}[b]
\begin{center}
\includegraphics[width=8cm]{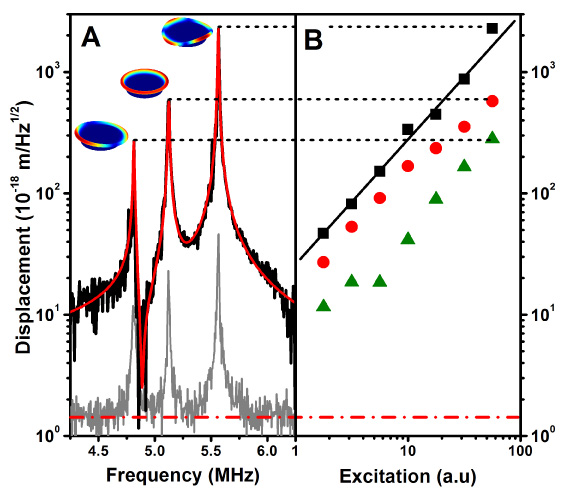}
\caption{(color online) Gradient force actuation of a COEMS. {\bf A} Square root transduction spectra. Black curve: measured spectra; red curve: theoretical model including interference between the mechanical modes; grey curve: Brownian motion spectra; dash-dotted line: transduction sensitivity. Insets: (from left to right) finite element models of the lowest order crown ($j=1$), lowest order radial flexural ($j \! = \! 2$), and second order crown ($j \! = \! 3$) modes. Resolution bandwidth: $\Gamma_{\rm RBW}/2\pi \! = \! 3$~kHz. {\bf B} Square root peak transduction spectra as a function of RF drive amplitude. Green triangles: $j \! = \! 1$, red circles: $j \! = \! 2$, black squares: $j \! = \! 3$.}
\label{spectra}
\end{center}
\end{figure}

The transduction spectrum obtained by spectral analysis of the in-loop transduction signal without gradient force actuation or feedback is shown by the grey curve in Fig.~\ref{spectra}A for 1~mW of incident probe power. The absolute mechanical displacement amplitude was calibrated by observing the response of the optical transmission to a known reference phase modulation\cite{Schliesser08}. One observes three characteristic spectral peaks due to Brownian motion of microtoroid mechanical modes. Finite-element modeling identified these modes to be the lowest order flexural mode, and the two lowest order crown modes; with the peak frequencies agreeing with the model to within 5\%.

To observe the quantum behavior of a mechanical oscillator  the transduction sensitivity must be sufficient to resolve the oscillators zero-point motion. Both the transduction sensitivity and the amplitude of zero-point motion of our COEMS can be determined from a fit to the transduced spectral density $S(\omega) \! = \! \sum_{j=1}^3{S_x^{(j)} (\omega)} \! + \! S_{N}$; where $S_{N}$ is the transduction noise due in our case to laser phase noise, and $S_x^{(j)} \! = \! 2 k_B T \Gamma_j m_j|\chi_j(\omega)|^2$ is the spectral density of Brownian motion of mechanical mode $j$\cite{Cohadon} with effective mass  $m_j$, damping rate $\Gamma_j$, and mechanical susceptibility $\chi_j(\omega) \! = \! [m_{j} (\omega_{m,j}^2 \! - \! \omega^2 \! - \! i \Gamma_j \omega )]^{-1}$. A transduction sensitivity of $S_N^{1/2} \! = \!  1.5 \! \times \! 10^{-18}$~m~Hz$^{-1/2}$ was established from the fit, three orders of magnitude better than any NEMS to date\cite{Knobel}; and the mode effective masses and damping rates were $(m_1,m_2,m_3) \! = \! (280,410,33)~\mu$g and $(\Gamma_1,\Gamma_2,\Gamma_3)/2\pi \! = \! (9.5,11.5,6.8)$~kHz, respectively. The peak of the zero point motion spectral density can then be calculated as $S_{zp}^{(j)} \! = \! \hbar/m_j \Gamma_j \omega_{m,j}$\cite{Schliesser08}, to give $({S_{zp}^{(1)}},{S_{zp}^{(2)}},{S_{zp}^{(3)}})^{1/2} \! = \! (1.4, 1.1, 4.6) \! \times \! 10^{-20}$~m~Hz$^{-1/2}$ for the three modes. The transduction sensitivity is, therefore, within two orders of magnitude of the mechanical zero-point motion. Techniques have recently been developed to substantially improve both the microtoroid transduction sensitivity through improved optical quality and shot noise limited homodyne detection\cite{Schliesser08}; and the mechanical mode damping rate and effective mass through nanofabrication\cite{Anetsberger}, cryogenic cooling\cite{Kippenberg09}, and evacuation of the media surrounding the microtoroid\cite{Anetsberger}. These techniques are fully compatible with the gradient force actuation scheme demonstrated here, and should enable sub-zero point motion transduction sensitivity in our COEMS architecture.

Gradient force actuation of the COEMS was characterized by applying the output voltage from a network analyzer to the sharp electrode, and monitoring the frequency response of the in-loop transduction signal. Fig.~\ref{spectra}A shows the observed mechanical frequency response for a $3$~V$_{\rm rms}$ network analyzer output voltage. A large increase in mechanical oscillation is observed when the RF drive frequency matches a mechanical resonance frequency. The maximum oscillation amplitude observed was $S_{x, {\rm max}}^{1/2} \! = \! 2.4 \! \times \! 10^{-14}$~m~Hz$^{-1/2}$ at the peak of the second order crown mode ($j \! = \! 3$), with an applied voltage of only 3~V$_{\rm rms}$. This corresponds to a peak-to-peak gradient force of  $F \! = \! 4 \pi^{-1/2} m_3 \omega_{m,3} \Gamma_{3}  \Gamma_{\rm RBW}^{1/2} S_{x, {\rm max}}^{1/2} \! = \! 0.40~\mu$N, surpassing all cryogenic COMS to date by more than an order of magnitude\cite{Kippenberg09}. The peak oscillation amplitude of each mechanical mode was found to be linear as a function of applied voltage, as shown in Fig.~\ref{spectra}B. A linear dependence on DC voltage applied to the sharp electrode was also observed, confirming gradient forces as the actuation mechanism\cite{Unterreithmeier09}.
Neither the presence of the sharp electrode, nor the RF drive, caused any observable degradation of mechanical quality factor; demonstrating the low-dissipation nature of the actuation.

The criterion $T \!\! < \!\! \hbar \omega_m/k_B$ must be met for the quantum behavior of a mechanical oscillator to dominate classical thermal fluctuations. For typical mechanical resonance frequencies this imposes the stringent condition of milli- to micro-Kelvin temperatures. In COMS, heating via optical absorption is a key concern, placing an upper limit on the sustainable intracavity power\cite{Kippenberg09}. Using radiation pressure, an intracavity power of $P\!\approx\!c F/\pi\!=\!36$~W would be required to achieve the maximum actuation force observed here\cite{Carmon05}. However, for microtoroids in a cryogenic environment Schliesser {\it et al.}\cite{Kippenberg09} found that just 1~W of intracavity power caused a 10~K temperature increase, precluding operation in the quantum regime. Hence, the COEMS presented here provides both the unique capacity to strongly drive mechanical oscillators in the quantum regime; and, transduction sensitivity close to the mechanical zero-point motion. Combined, these attributes provide new quantum control capabilities for mechanical systems, as well as an enabling step towards the observation of mechanical quantum nonlinear dynamics.

To demonstrate electro-optic feedback cooling we introduce a feedback loop using gradient force actuation to apply the in-loop transduction signal back upon the mechanical motion.
Delaying the feedback by a quarter cycle provides a viscous damping force, which both cools and damps the mechanical motion\cite{Cohadon}. Including spectrally flat in-loop transduction noise,  the final temperature $T$ of the oscillator under feedback with gain $g$ can be shown to be\cite{Poggio}
\begin{equation}
\frac{T}{T_{0}} = \left [1 + \frac{g^2}{\rm SNR} \right ] \times \frac{1}{1+g}, \label{temp}
\end{equation}
where $T_0$ is the initial temperature, and ${\rm SNR} \! = \!  S_{x,0} (\omega_m)/ S_N^{\rm IL}$ is the signal-to-noise ratio of the peak of the in-loop transduction spectra without feedback to the in-loop transduction noise.
One sees that feedback induces cooling. However, competing heating due to transduction noise imprinted on the mechanical oscillator imposes a minimum achievable temperature of $T_{\rm min} \! = \! 2 T_0 [ \sqrt{1 \! + \! {\rm SNR}}-1 ]/{\rm SNR}$ at $g \! = \! \sqrt{1 \! + \! {\rm SNR}}-1$, with the temperature increasing at higher gain.

\begin{figure}[b!]
\begin{center}
\includegraphics[width=8cm]{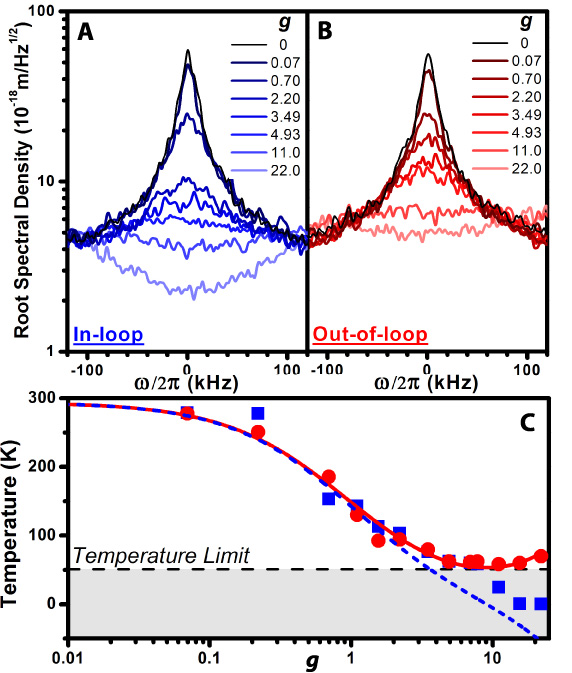}
\caption{(color online) Feedback cooling with varying feedback gain. {\bf A} and {\bf B}: In-loop and out-of-loop transduction spectra.
{\bf C}: Temperature inferred using in- and out-of-loop transduction signals. The red circles \textcolor{red}{$\medbullet$} and blue squares \textcolor{blue}{$\blacksquare$} respectively denote  out- and in-loop temperature inferences; the solid red curve and dashed blue curve respectively denote the theoretical predictions of {\it actual} mechanical oscillator temperature and inferred in-loop temperature\cite{note}.}
\label{feedbackcooling}
\end{center}
\end{figure}
Figure~\ref{feedbackcooling}A~and~B respectively show the effect of the feedback on the in- and out-of-loop transduction spectra for a radial flexural mode at $6.272$~MHz with an effective mass of $30 \! \pm \! 10$~$\mu$g and $11.5$~kHz damping rate. In both cases a significant reduction in  mechanical noise power, and hence cooling, is observed with increasing feedback gain. However, a rapid acceleration of apparent cooling is seen via in-loop transduction for gains $g \! > \! 3.5$ with eventual inversion of the mechanical response at $g \! > \! 10$; in stark contrast to the observations from out-of-loop transduction. This dramatic {\it squashing} of the noise spectra to below the transduction noise level is the result of feedback induced anti-correlations between the in-loop transduction noise and mechanical oscillator motion\cite{Cohadon,Poggio,Schwab}.

When mechanical motion and transduction noise are uncorrelated, the mode temperature is proportional to the integrated area between the transduction spectra and the transduction noise\cite{Poggio}. Fig.~\ref{feedbackcooling}C shows the in- and out-of-loop mechanical oscillator temperatures inferred
in this way as a function of feedback gain.  The out-of-loop temperature is in good agreement with the theoretical prediction of Eq.~(\ref{temp}). A minimum temperature of $T \! = \! 58~{\rm K} \! > \! T_{\rm min} \! = \! 53$~K was observed for $g \! = \! 8$, limited by the in-loop transduction noise which gave a signal-to-noise ratio of ${\rm SNR} \! = \! 100$; with the temperature, as predicted, observed to {\it increase}  at higher gains due to feedback noise imprinted on the mechanical oscillator. In contrast, at high gain anti-correlations between oscillator motion and transduction noise cause the in-loop temperature inference to diverge significantly from theory, dropping well below the theoretical limit and passing through $0$~K before becoming unphysical on inversion of the observed mechanical response. The result is a paradoxical and erroneous continual {\it reduction} in the in-loop temperature inference with increasing gain. These results dramatically demonstrate the requirement of independent temperature verification in feedback cooling, as first demonstrated here. Independent verification is essential, not only in the regime of high feedback gain, but also in the critical quantum paradigm where measurement backaction itself perturbs and correlates the mechanical oscillator and transducer.

Although this proof-of-principle demonstration achieved only a modest final phonon occupation number of $\langle n \rangle \! = \! k_B T/ \hbar \omega_m \! = \! 19,000$, the technique is fully compatible with the recent advancements in microtoroid opto-mechanics discussed earlier\cite{Anetsberger,Kippenberg09,Schliesser08}. Implementing these techniques from a cryogenic starting temperature of 1.7~K, a near ground state final phonon occupation number of $\langle n \rangle \! \approx \! 0.7$ could be achieved\cite{Genes}.

We have demonstrated a cavity opto-electromechanical system which combines the ultrasensitive transduction capabilities of cavity optomechanics with gradient force control capabilities from nanoelectromechanics. Electrical gradient forces as large as $0.40~\mu$N were achieved, significantly higher than has been demonstrated with radiation pressure actuation without substantial heating.
Simultaneously, a transduction sensitivity of $1.5 \! \times \! 10^{-18}$~m~Hz$^{-1/2}$ was observed, less than two orders of magnitude away from the mechanical zero-point motion. Electrically actuated, optically transduced feedback cooling was achieved for the first time as a demonstration of the control capabilities of our system. The implementation of an out-of-loop probe allowed the first independent temperature verification, illustrating both the transduction noise limit of feedback cooling and the striking effect of squashing on in-loop temperature inferences.
Our results represent important progress in the control of mechanical systems at the quantum level; as well an enabling step towards the new regime of quantum nonlinear mechanics, where strong mechanical driving of a ground-state cooled mechanical oscillator allows exploration of nonlinear quantum dynamics.

This research was funded by the Australian Research Council grant DP0987146, and performed in part at the Australian National Fabrication Facility. We gratefully acknowledge helpful advice  from T. Kippenberg, G. Milburn, and T. Stace; and experimental support from A. Rakic and K. Bertling.

\end{document}